# The sub-mm J=6-5 line of $^{13}$CO in Orion

T. L. Wilson
*Naval Research Laboratory, Code 7210, Washington, DC 20375*
*tom.wilson@nrl.navy.mil*

D. Muders
*Max-Planck-Institut f. Radioastronomie, Auf dem Huegel, 53121 Bonn, Germany*

M. Dumke
*European Southern Observatory, Alonso de Córdova 3107, Vitacura, Santiago, Chile*

C. Henkel
*Max-Planck-Institut f. Radioastronomie, Postfach 2024, 53121 Bonn, Germany*

Jonathan H. Kawamura
*Jet Propulsion Laboratory, California Institute of Technology, Pasadena, CA 91109*

**ABSTRACT**

We present a fully sampled map covering the Orion Hot Core and dense molecular ridge, in the sub-millimeter J=6-5 rotational transition of $^{13}$CO, at λ=0.45 mm with a resolution of 13" and 0.5 km s$^{-1}$. The map covers 3' by 2'. The profile centered on the Hot Core peaks at 8.5 km s$^{-1}$ and has a peak intensity of 40 K, corrected antenna temperature. It shows line wings from 30 km s$^{-1}$ to -20 km s$^{-1}$. The map of intensity, integrated from 0 to +18 km s$^{-1}$, shows a prominent maximum <5" from the center of the Orion Hot Core. The FWHP is 37", larger than the regions containing complex molecules. Single dish measurements of lines from the J=2-1 or J=1-0 transitions of CO isotopes show no such distinct maximum. Correcting for τ=1.5 in the J=6-5 line of $^{13}$CO, and assuming that the level populations are thermalized at 150 K, the beam averaged column density between 0 to +18 km s$^{-1}$ is N($^{13}$CO)=6.8 · 10$^{17}$ cm$^{-2}$ and N(CO)=5.2 · 10$^{19}$ cm$^{-2}$. When combined with published dust emission data, the CO/ H$_2$ number ratio is 2 · 10$^{-5}$, a factor of ~5 lower than the canonical value, 10$^{-4}$. For the Orion South and Orion Ridge region, the column density of CO is <25% of that found for the Hot Core but CO/H$_2$ ratios are similar. Models of Photodissociation Regions, PDRs, predict that CO lines from PDRs are only marginally optically thick. Thus our map traces warm and dense molecular gas rather than PDRs.

*Subject headings:* ISM: individual (Orion BN/KL, Orion Hot Core) - ISM: jets and outflows - ISM: Molecules-stars: early-type- stars: submillimeter

## 1. Introduction

Although molecular clouds are composed mostly of $H_2$, this species has no allowed transitions and usually has no populated energy levels close to the ground state, so the usual approach used in cloud mapping is to measure rotational transitions of carbon monoxide (CO) and from these data deduce the beam averaged column density and density of $H_2$. The J=1-0 transition is well-suited to determine the boundaries and gross properties of molecular clouds (Dame et al. 2001). Stars are formed in high density regions (see, e.g., Larson 2003), where complex interstellar chemistry reactions take place (see, e.g., Herbst & van Dishoeck 2009). High mass stars are found in warm and dense regions, so measurements of emission from higher rotational levels of CO are well suited to such more detailed studies; in addition, the population of vibrationally excited states of CO is small. In lower density clouds Frerking et al. 1982 found a $CO/H_2$ number ratio of $10^{-4}$. In cold, very dense regions, the CO can condense onto grains, so the ratio is measured to be significantly lower (Alonso-Albi et al. 2010). In hot, high excitation regions, the abundance may be lowered if much of the atomic oxygen is incorporated in water vapor (see Haranda et al. 2009).Thus the CO rotational transitions are often optically thick, so detailed estimates of column densities are better carried out with the use of rarer isotopes, such as $^{13}C^{16}O$ (hereafter $^{13}CO$) and then multiplying this result by the isotopic ratio. Studies making use of rotational transitions of CO isotopes have the advantage of: (1) large abundance, (2) excitation which is close to Local Thermodynamic Equilibrium (LTE) and (3) simpler interstellar chemistry, since the CO molecule has a high dissociation energy and is rather stable so will not be easily converted to other species, at least in cold molecular clouds. The higher rotational transitions of the CO molecule require both larger $H_2$ densities and kinetic temperatures, so the emission from such transitions arises only from high excitation regions. Other species can mark out regions of high density or kinetic temperature, but are less abundant or are subject to more complex chemistry or excitation.

The Orion complex is among the most studied sources in our galaxy. At a distance of ~420 pc (see, e. g., Sandstrom et al., 2007, Menten et al. 2007; Hirota et al. 2007), this is the nearest site of recent high mass star formation. The prominent HII regions NGC 1976 or M42 and NGC 1977 are on the near side of a large molecular cloud associated with Lynds dark cloud L1640 (see, e.g. O'Dell 2001). To the rear of the HII region is the Orion Molecular Cloud, OMC-1. The interaction of the ionized and molecular gas has an important effect on both the HII region and the molecular cloud (see, e.g., van der Werf & Goss 1989). The map of Maddalena et al. (1986) in the J=1-0 transition of CO shows that the OMC-1 cloud has an irregular, streamer-like shape, in the form of an integral-sign, with an extent of ~10$^o$. The brightest molecular emission in OMC-1 is near the HII region NGC 1976 (see, e.g., Tatematsu et al. 1993). Plume et al. (2000) measured the J=5-4 transition of $^{13}$CO over a 0.5$^o$ by 2$^o$ area with a 3.2' resolution. The brightest emission has an extent of ~15' in the north-south by ~5' in the east-west direction. Within this are two prominent maxima, the Kleinmann-Low, KL, infrared nebula (see, e.g., Gezari 1992, Dougados et al. 1993) and Orion South (Zapata et al. 2005). Both are centers of emission from complex molecules, masers, and outflows. In molecular lines, the KL nebula is referred to as the Hot Core (see, e.g., Wright et al. 1996, Blake et al. 1996, Wilson et al. 2000, Guélin et al. 2008). In the image of Guélin et al. (2008) methyl formate in the Hot Core is seen as a V-shaped region with arms 12" long and 3" wide, oriented 18$^o$ E of N with an opening angle of 38$^o$. This region shows large extinction even in the near IR. In addition to extended IR emission from warm dust, there are ~20 near-IR sources found toward the Hot Core (Dougados et al. 1993). Because of crowding, sub-arc second angular resolution combined with accurate astrometry is needed for secure identifications. The source IRc2 had been identified as the driver of the high velocity CO, SiO and H$_2$O outflows up to ~100 km s$^{-1}$ (see, e.g., Genzel & Stutzki 1989). Recently, more accurate positional data showed that the situation is more complex (Gezari 1992, Dougados et al. 1993, Gezari et al. 1998). With more recent high angular resolution radio data, the location of the outflow driver has been identified with the highly obscured source "I" (Menten & Reid 1993), but an additional outflow source in the Hot Core has been identified by de Vicente et al. (2002) and Beuther & Nissen (2008). Both the Becklin-Neugebauer (BN) and I point continuum sources show large proper motions (Rodríguez et al. 2005). In our 13" beam, these positions cannot be distinguished, so we have chosen the nominal position of IRc2 as our map zero point; we will refer to this as IRc2/I (see Gaume et al. 1998, figures 2, 3 and 4).

The stars ionizing NGC 1976 may contribute to the heating of the Hot Core. On the surfaces of dense molecular clouds, ultraviolet radiation, from nearby O and B stars, causes a transition from molecular to atomic and ionized gas, giving rise to Photon Dominated Regions or PDRs (see Hollenbach & Tielens 1997). In those PDRs, some molecular species are present, although the abundances are different, usually lower, than those in molecular clouds. PDRs have a high abundance of atomic oxygen, OI, and ionized carbon, CII, but OI is also found in molecular clouds (Caux et al. 2001). Within the Orion complex, the Orion Bar, located ~3.6' SW of the center of NGC 1976, is a prominent PDR. From the geometry of the Orion Bar, the transition from ionized to atomic and molecular components can be separated in angle (see, e.g., Hogerheijde et al. 1995, Leurini et al. 2010, van der Wiel et al. 2009). There is no such angular separation for the interface between NGC 1976 and OMC-1. The CN molecule is a tracer of high density PDRs (Rodríguez-Franco et al. 2001). For the NGC 1976-OMC-1 interface, measurements of mm and sub-mm spectral line emission from the CN molecule were combined to show that this PDR is a thin, hot dense layer of material at the surface of the molecular cloud.

To further investigate the physical state of extended, warm gas, we have carried out measurements of the J=6-5 rotational transition of $^{13}$CO. This line arises from an energy level 111 Kelvin above the ground state and requires a density of $\sim 2 \cdot 10^5$ cm$^{-3}$ for excitation.

## 2. Observations

The J=6-5 line $^{13}$CO data were taken with the 10 meter Heinrich Hertz Telescope (HHT) on Mount Graham AZ in February 2002. For a description of this instrument, see Wilson et al. (2001a). At the line frequency, 661.067270 GHz (Lovas 1992), the FWHP beam size of the HHT is 13". The forward and the main beam efficiencies are $\eta_f$ =0.95 (for sources of size ~30') and $\eta_B$=0.4 (for sources comparable to the beamsize), respectively (Wilson et al. 2001a). The calibration was carried out using the chopper wheel method. No corrections for forward or beam efficiency were made, so the temperature scale in our figures is very nearly $T_A^*$. Thus these temperatures are corrected for absorption by the earth's atmosphere and are appropriate for very extended emission regions. For compact sources, we divided the main beam efficiency by $\eta_B$ to convert $T_A^*$ to $T_{MB}$. The pointing was checked by continuum measurements of Saturn, and found to be accurate to 2", RMS.

The front end was a superconducting hot-electron bolometer mixer provided by the Harvard-Smithsonian Center for Astrophysics (Kawamura et al. 2000). The spectra were analyzed using an Acoustic Optical Spectrometer (AOS). The frequency resolution was 1 MHz, but all spectra were Hanning smoothed to a resolution of 2 MHz or 0.96 km s$^{-1}$. The receiver single sideband noise temperature, relevant for spectral line measurements, was 1800 K. The atmospheric optical depth in the zenith was of order unity. The single sideband system temperature, corrected for loss in the earth's atmosphere was 14000 K.

The spectral line data were taken with on-the-fly mapping. The total map size is 3' (north-south) by 2' (east-west), including the Hot Core and Orion South. The reference position was chosen to be 10' to the west of the central position. This location has only weak CO emission in the higher rotational transitions (Howe et al. 1993). The observing method consisted of a calibration, and then an off-source spectrum followed by a row of on-source spectra. The final spectra are the ratio of on minus off divided by off. The integration time for each final spectrum was 10 seconds. The exact number of on-source spectra for each comparison spectrum depended on the frequency baseline response.

In the frequency baselines of the final spectra, there were instrumental effects, ripples, caused by changes in weather or reflections in the telescope between off and on-source measurements. These ripples were removed using the Fast Fourier Transform algorithm in the CLASS software package. After this, a low order baseline was removed. The final spectra had baselines dominated by noise. After baseline removal, the spectra were regridded on a rectangular grid in $(\alpha, \delta)$ with a separation of 5". The typical RMS noise in an individual on-off spectrum is 3.0 K.

## 3. Results

In Figure 1 we show a comparison of spectra taken at the (0, 0) offset, the position of the source IRc2/I. The CO J=7-6 spectrum is above (Wilson et al. 2001b) and the $^{13}$CO J=6-5 spectrum is the histogram below. The $^{13}$CO line profile shows an excellent agreement with the result of Schilke et al. (2002, taken with the CSO 10-m having the same beamsize). The velocities are with respect to the local standard of rest, using the standard solar motion. The region between the two vertical lines, at 0 and 18 km s$^{-1}$, are meant to include gas not affected by an outflow. Since emission from the Hot Core is rather compact, the correction for main beam efficiency to the CO J=7-6 line raises the peak intensity to $T_{MB}$=200 K, main beam brightness temperature. Some of the intensity is from the extended Orion Ridge, so we have assigned $T_K=T_{MB}$=150 K to the Hot Core. There are significant differences between the CO and $^{13}$CO profiles in figure 1. In the CO J=7-6 profile, the line wings extending to

$V_{LSR}$ >|80| km s$^{-1}$, arise from the high velocity outflow. In $^{13}$CO, these wings are not detectable for

> $|30|$ km s$^{-1}$. At higher V$_{LSR}$, the lower intensity of the $^{13}$CO indicates that the CO line is becoming optically thin, so the intensity of the $^{13}$CO emission approaches 1/76 that of the CO (see Stahl et al. 2008), too weak for detection. We have produced maps for velocity ranges of the $^{13}$CO J=6-5 emission, from 15.1 to 40 km s$^{-1}$ and 2.3 to -20 km s$^{-1}$, to study the lower velocity portions of the outflow. From these maps, the velocity features are centered on the (0", 0") position. In the J=7-6 CO outflow, the higher positive velocity line wings are 20" north and the negative velocity wings are 20" west of the (0", 0") position (see Wilson et al. 2001, Furuya & Shinnaga 2009). For more quiescent emission, the shapes of the $^{13}$CO and CO lines agree fairly well, if the notch at 4 km s$^{-1}$ in the CO spectrum may be caused by emission in the reference spectrum (see Howe et al. 1996).

The $^{13}$CO profiles in Figure 2 show that at offset positions, the lines have narrower widths, and somewhat lower peak temperatures. The spectra in panels (a) and (b) are from Orion Ridge emission at (0", -40") and (0, -55"). These are rather similar. The (0", -40") offset in panel (a) is a position in the Orion Ridge where no outflows are present. The (0, -55") offset in panel (b) is at the Declination of the brightest Trapezium star, $\theta^1$C Orionis. The profile for offset (-24", -100") in panel (c) is for the position of Orion South. This line shows a wing toward negative velocities; this source has water masers and a highly collimated outflow (Gaume et al. 1998, Zapata et al. 2010). The profile in panel (d), at an offset (126", -146"), was taken toward the Orion Bar (See Hogerheidje et al. 1995).

In Figure 3 we show the emission from individual velocity channels from 5.8 to 12.8 km s$^{-1}$. These velocities cover the range of lower velocity emission from the Hot Core and Orion Ridge, but little from the high velocity CO outflow. As in measurements of ammonia (see, e.g., Cesaroni & Wilson 1994), there is lower velocity extended emission at 7.62 and 8.49 km s$^{-1}$ to the south and weak emission at 9.36 and 10.23 km s$^{-1}$ to the north of the Hot Core. We identify this lower intensity extended emission with the Orion Ridge.

In Figure 4 we show the emission integrated from 0 to 18 km s$^{-1}$. The largest peak is located at (-6", -7") from the map center. Given pointing uncertainties and source size, the position of our J=6-5 maximum is consistent with being centered on IRc2/I, which is near the eastern boundary of the Hot Core. Emission from Orion South and the Orion Ridge is more restricted in velocity than the Hot Core. This is confirmed by line profiles (Figure 2) taken at these positions. At the lowest contour, the integrated emission covers a region extending over 1.5' in α by 2.5' in δ at the level of

50 K·km s$^{-1}$. The weaker peak, at an offset of (-24", -100"), is Orion South which is an ammonia, collimated CO outflow and water maser source (see, e.g., Zapata et al. 2010, Gaume et al. 1998). We identify the extended lower intensity emission between the Hot Core and Orion South with the Orion Ridge. This has a maximum value of 100 K·km s$^{-1}$, $T_A^*$. Subtracting this value from the measured peak intensity of the Hot Core emission leaves 500 K·km s$^{-1}$, $T_A^*$ or 1250 K·km s$^{-1}$, $T_{MB}$. The deconvolved FWHP sizes of the Hot Core are 36" in α by 38" in δ. The Hot Core emission seems to show a slight extension to the northeast; the dust emission maps show a second peak in this direction (see, e.g., Mezger et al. 1990). At 420 pc, the geometric mean of these angular sizes is equivalent to a FWHP size of 0.07 pc. The $^{13}$CO FWHP size of the Hot Core is much larger than the *total* size, 0.02 pc, occupied by complex molecules in the Hot Core (see, e.g., Figure 4 of Guélin et al. 2008). This is to be expected, since complex molecules are fragile, compared to CO which has a dissociation energy of 11 eV (see, e.g., Lequeux 2004).

Figure 5 shows the ratio of the J=6-5 $^{13}$CO to the J=7-6 CO emission, both integrated from 0 to 18 km s$^{-1}$. The ratio reaches a maximum of 0.6 at the Hot Core, and a minimum of 0.3 elsewhere. The extension seen in J=6-5 $^{13}$CO to the NE is close to the noise level but extended so is significant. From this map, there is significant emission from J=6-5 $^{13}$CO in the Hot Core; elsewhere there is less J=6-5 $^{13}$CO emission. The measured ratio should be multiplied by 76 the local $^{13}$C/$^{12}$C ratio (Stahl et al. 2008) to determine the CO optical depth, so the optical depth for the Hot Core is larger than 20 in the J=7-6 line of CO. It may be significant that at Δδ=-55" there is a minimum in the ratio, since this is the Declination of the brightest Trapezium star, $\theta^1$C Orionis. However, as will be discussed later, it is not possible to separate CO emission from the dense molecular gas from PDR emission on the basis of the results in Figure 5.

## 4. Comparison with other results and analysis

### 4.1 Other CO maps

Marrone et al. (2004) measured a 4' by 6' region in the CO J=9-8 line with an 84" beam. The emission is present at all positions in their map, supporting their interpretation as CO line emission from the PDR. The map of the J=7-6 line of CO with a 13" resolution was presented by Wilson et al. (2001b). In their Figure 2, velocity channel maps from 1 to +16 km s$^{-1}$ show a maximum ~10" east of the IRc2/I position with a lower intensity extension to the north. The Orion South source is not prominent. Images of the J=1-0 line of $^{13}$CO made with the IRAM 30-m telescope have an angular resolution of 23" (Mauersberger et al. 1989), who found extended emission but no compact peak at the location of the Hot Core. A comparison with our Figure 4 shows that this cannot be caused by the larger telescope beam since the Hot Core has a FWHP size of 37" in the J=6-5 line of $^{13}$CO, so must be due to excitation. The image of Plume et al. (2000) of $^{13}$CO J=5-4 was made with a 3.2' resolution, covering ~2 degrees NS by 30' EW. The distribution of integrated intensities (their Figure 1) shows a region extended over 10' NS including the Hot Core. Plume et al. (2000) found an average H$_2$ density of $10^4$ cm$^{-3}$, a $^{13}$CO column density of $2 \cdot 10^{16}$ cm$^{-2}$ and kinetic temperatures, $T_K$, in the range 20 K to 45 K. Thus, this must be emission from the Orion ridge. In the data of Plume et al. (2000) there is no prominent peak at the Hot Core. This is probably due to dilution in their 3.2' beam. Compared to the Orion ridge, the Hot Core is more prominent in the J-6-5 line of $^{13}$CO, so the excitation of this molecule is higher in the Hot Core. The column density of CO and kinetic temperature in the Hot Core must exceed that in the Orion ridge.

### 4.2 Analysis of our results

The integrated, optically thin, line intensity from Eq. (15.26) of Wilson et al. (2008) gives the beam averaged column density,

$$N(^{13}\text{CO}, J = 5) = 3.2 \cdot 10^{13} \int T_{MB}\, dv \qquad (1)$$

The optical depth of the $^{13}$CO J=6-5 line could be significant. We can estimate this from the data of Schilke et al. (2001). For the solar system, the ($^{12}$C/$^{13}$C)($^{16}$O/$^{18}$O) ratio is 5.5 (see, e.g. Wilson & Rood 1994). If both C$^{18}$O and $^{13}$CO are thermalized at the same temperature, we can use Eq. (15.38) of Wilson et al. (2008) to determine the optical depth of the J=6-5 line. This is τ=1.5. Using Eq. (15.26) of Wilson et al. (2008), we find that the correction for optical depth is 2.0. The measured column density in the J=5 level, corrected for the integrated intensity using a beam efficiency $\eta_B$=0.4 gives a value $T_{MB}\, \Delta v$=1.2 $10^3$ K·km s$^{-1}$. Combining the optical depth correction with the measured $T_{MB}\, \Delta v$, the beam averaged column density in the J=5 level of $^{13}$CO is $7.6 \cdot 10^{16}$ cm$^{-2}$. For the calculation of the total population, we assume that the CO energy levels are populated in LTE at 150 K. Using Eq. (15.34) and (15.35) of Wilson et al. (2008), we obtain the relation between the total column density and that in the J=5 level:

$$N(^{13}CO) = \frac{1}{2J+1} \cdot \frac{kT_K}{hB_e} \cdot N(^{13}CO, J=5) \cdot e^{\frac{79}{T_K}} \qquad (2)$$

where 2J+1 is the statistical weight, and $B_e$ is the $^{13}CO$ rotational constant expressed in Kelvins. Using these values, we obtain $N(^{13}CO)=6.8 \cdot 10^{17}$ cm$^{-2}$. If we assume that the CO/$^{13}CO$ ratio is 76, the value measured for the local ISM (Stahl et al. 2008), the total beam averaged column density of CO is $5.2 \cdot 10^{19}$ cm$^{-2}$. This is given in Table 1.

Another approach is to apply the Large Velocity Gradient (LVG). With only a few $^{13}CO$ lines, this leads to only approximate values. The upper limit to densities cannot be more than ten times the critical density for the J=6-5 transition, which is $n^*=2.8 \cdot 10^5$ cm$^{-3}$. Our data are consistent with a density $n(H_2)=5 \cdot 10^6$ cm$^{-3}$ and kinetic temperature 150 K (see, e.g., the values in Blake et al. 1996). The LVG approximation also provides an estimate of the CO/$H_2$ ratio per velocity gradient, in units of km s$^{-1}$ pc$^{-1}$. If we estimate this gradient as the ratio of observed linewidth to source size, the value is 40 km s$^{-1}$ pc$^{-1}$, and the CO/$H_2$ ratio is $3 \cdot 10^{-5}$, less than the canonical value, $10^{-4}$ (Frerking et al. 1982). The uncertainties in all of these estimates could be a factor of two.

The $H_2$ densities and kinetic temperatures in the more extended Orion Ridge must be lower, since for single dish measurements, the ridge dominates the CO emission from the region toward and around the Hot Core position in the CO and $^{13}CO$ J=2-1 and J=1-0 lines, but not the J=6-5 or J=7-6 lines. Repeating the calculation above, but with a kinetic temperature of 50 K, we obtain a total CO column density of $4.6 \cdot 10^{18}$ cm$^{-2}$. Another value can be estimated from the LVG approximation. This gives an $H_2$ density of $\sim 10^5$ cm$^{-3}$, for a kinetic temperature 50 K. The values are somewhat larger than those given by Plume et al. (2000), namely $10^4$ cm$^{-3}$ and 45 K.

For the position of Orion South, the integrated emission in our map is 100 K $\cdot$ km s$^{-1}$, $T_{MB}$. Assuming that the CO population is thermalized at the kinetic temperature, 100 K, we obtain a total CO column density of $4.3 \cdot 10^{18}$ cm$^{-2}$.

**4.3 Comparison with dust emission maps**

The column density of $H_2$ can be estimated from measurements of millimeter and sub-millimeter continuum emission provided this emission is from dust grains. We make use of the bolometer map of Mezger et al. (1990) that had an 11" resolution, nearly the same as used in our map. The center positions of the dust and $^{13}CO$ agree rather well, but the angular size of the Hot Core from dust emission is ~50% of the size from $^{13}CO$. This may be due in part to the beam-switched data taking procedure used for dust emission. The separation of the dust emission from the Hot Core from the Orion Ridge is a subject of interpretation. We estimated that the Hot Core dust emission has a flux density of 20 Jy. The emission is more prominent in dust emission than J=6-5 $^{13}CO$. In addition the emission from Orion South has $S_\nu$=7.7 Jy and the offset at (0", -40") in the Orion Ridge has 2.3 Jy. To obtain a CO/ $H_2$ ratio, we require additional assumptions about grain properties. For the simplest situation, one can apply Eq. (10.7) of Wilson et al. (2008)

$$N(H) = 1.93 \cdot 10^{24} \frac{S_\nu}{\theta^2} \cdot \frac{\lambda^4}{z/z_\odot \, b \, T_D} \qquad (3)$$

The $H_2$ abundance is ½ of the H abundance, the wavelength $\lambda$ is in mm, the flux density, $S_\nu$, in mJy and the FWHP beamsize, $\theta$, is in arc seconds. We have set the adjustable parameter "b" to unity and assumed that the metallicity is the solar value. At densities $>10^5$ cm$^{-3}$, the dust and gas temperatures are equal. For these regions, we assume that the dust temperature, $T_D$, equals the kinetic temperature, $T_K$. We collect these results in Table 1.

For both the ridge and the source Orion South, the dust emission is about a factor of two weaker than from the Hot Core, but since the dust temperature is assumed to be one half of that for the Hot Core, $H_2$ column densities are nearly the same. The CO/$H_2$ ratio for Orion South is $2 \cdot 10^{-6}$, so for all three regions, the CO/$H_2$ ratios are smaller than the usually accepted value of CO/ $H_2=10^{-4}$ (Frerking et al. 1982). Since the flux densities of the dust emission were measured using beam switching, the actual values may be larger, so the CO/ $H_2$ values may be even lower. There are a few approaches that could raise the CO/$H_2$ ratio. The separation of Hot Core dust emission from the more extended Orion Ridge emission will affect our estimates of the $H_2$ column densities by a factor of two, but would not raise the $CO/H_2$ ratio significantly. If the temperatures of the dust and CO rotational levels are raised from 150 K to 250 K, the ratio will increase the $CO/H_2$ ratio by a factor of 2.8, because the total population of CO levels is proportional to the rotational temperature, while the column density of $H_2$ molecules is inversely proportional to the dust temperature. Thus, if one increases these temperatures, one can increase the $CO/H_2$ ratio. However, there is little evidence at present for large areas in the Hot Core with $T_K$ above 250 K. The temperatures in the Orion ridge are even lower, and there, the $CO/H_2$ ratio is smaller. Thus, varying dust and CO rotational temperatures will not raise the $CO/H_2$ to any large extent. In addition our estimates of the temperatures appear to be realistic. The ratio estimated from the LVG approximation is also smaller than the usual value, so it would appear that our estimate of the $CO/H_2$ ratio is smaller than the canonical value, $10^{-4}$, from both the LVG analysis and this comparison with dust emission results. The ratio in the Orion Ridge is also low. The reason for this result cannot be freezing out onto cold dust grains. It is also unlikely that the outflows and nearby high mass stars could destroy CO but not the complex molecules in these clouds. As Harada et al. (2010) point out, at high kinetic temperatures, the abundance of CO is lower since oxygen is combined in the form of water vapor. In their Figure 2, they show the abundance of CO as a function of time for the harsh environment of Active Galactic Nuclei. This may be similar to the Orion Hot Core. At short times, the $CO/H_2$ ratio is $\sim 10^{-7}$, rising at later times to a $CO/H_2$ ratio of $8 \cdot 10^{-5}$.

For the Hot Core, the $H_2$ column densities and FWHP size can be combined to give an average local density of $1.3 \cdot 10^7$ $cm^{-3}$ and mass of 73 solar masses. These values are within factors of 2-3 of estimates made by e.g. Mezger et al. (1990). The density is consistent with the value used to estimate the CO column density.

**4.4 Relation to Complex Molecules and Outflows in the Hot Core**

The total extent of the Hot Core in complex molecular species is less than 10" (from the data of Guelin et al. 2008). The absence of self absorption indicates that the excitation (densities and temperatures) is larger in the part of the Hot Core which is closer to NGC 1976. The IR sources, especially IRc2/I are known to have substantial extinction even at near IR wavelengths (see Dougados et al. 1993, Gezari et al. 1998), so these sources must be located toward the back of the Hot Core. The eastern part of the Hot Core "V" structure is warmer than the western part, so there must be more heating sources in or close to this feature (see, e.g. Gaume et al. 1998). In addition to heat sources, there are outflows. Our data is not sensitive to the high velocity outflow (see e.g., Beuther & Nissen 2008), so our 13" resolution results cannot be related to the energetic effects seen in $H_2$ (Stolovy et al. 1998), the motion of BN (Rodríguez et al. 2005) and the breakup of a star cluster (Zapata et al. 2009).

### 4.5 CO emission from the PDR

The brightest star of the Trapezium cluster is located at an offset of (+17", -47") from the center position of our map. Wilson et al. (2001b) had selected spectra along a line in $\alpha$ at $\Delta\delta$=-55" as the best location to find CO emission from the OMC-1 PDR. Along this offset, one expects that emission from the PDR would be most intense. From the PDR model of M. Wolfire (see Kaufman et al. 2006), the intensity ratio of the $^{13}$CO line to the CO line should be close to the ratio of $^{13}$C to $^{12}$C abundance, 1/76. From figure 5, the $^{13}$CO/CO ratio is 0.6 for the center of the Hot Core but 0.3 for $\Delta\delta$=-55", where the PDR should have the largest effect. It is not possible to separate the CO from the PDR from emission from dense molecular gas. However, as will be shown below although PDR emission contributes significantly to the CO J=7-6 emission, it does not contribute to the $^{13}$CO J=6-5 emission at the noise level of our map.

Our spectrum for the Orion Bar (Fig. 2) shows a rather intense and narrow $^{13}$CO J=6-5 line. This intensity is larger than one would predict if the $^{13}$CO were 1/76$^{th}$ of the CO line. For the Orion Bar, it is accepted that geometry plays an important role in determining the strength of this line. The Orion Bar is almost a sharp edge parallel to the line of sight (Hogerheijde et al. 1995, Lis et al. 1997); in contrast, most of the PDR between OMC-1 and NGC1976 is nearly face on. For a face-on geometry, we expect that the optical depth of all CO lines from the PDR between OMC-1 and NGC 1976 is much lower, so $^{13}$CO line emission from the PDR will be much weaker relative to the CO in the interface between NGC 1976 and OMC-1 than in the Orion Bar. Thus the intensity ratio of the CO emission from the Orion Bar is not a contradiction to the statement that the $^{13}$CO to CO intensity ratio in the OMC-1 PDR is close to the isotopic ratio.

From the multi-line analysis of Rodríguez-Franco et al. (2001), the CN emission traces the densest parts of the PDR. The peak of the CN emission is offset to the east of the Orion Ridge measured in $^{13}$CO J=5-4 (Plume et al. 2000) and our J=6-5 line. Rodríguez-Franco et al. (2001)

found an $H_2$ density $3 \cdot 10^6$ cm$^{-3}$ near the Trapezium and $10^5$ cm$^{-3}$ toward the Orion Ridge. When combined with PDR models, they deduced that the CN emission arises from a layer $10^{15}$ cm thick. Figure 1 of Rodríguez-Franco et al. (2001) shows that the distribution of CN has a width of $>10^{18}$ cm, so we must be observing the CN layer almost face-on with the dense Orion Ridge behind and offset to the west of NGC 1976. The radial velocities of the CN, the J=6-5 $^{13}$CO and J=7-6 transition of CO are similar, so the CN layer is at rest relative to the molecular cloud, in agreement with Rodríguez-Franco et al. (2001) . The PDR analysis of the fine structure IR lines of OI and CII (Herrmann et al. 1997) gave a density of $3 \cdot 10^5$ cm$^{-3}$ and $T_K$ of 300 K to 500 K, but much of the OI may arise in dense molecular clouds (Caux et al. 2001) so this analysis should be reconsidered. The 50" resolution CII line image of Herrmann et al. (1997) and the 84" data of Marrone et al. (2004) show the overall morphology of the less dense PDR, with the most intense emission to the west, north and south and weaker emission to the east of NGC 1976. Thus to the east, the HII region can expand more freely. This is reflected in the high resolution radio continuum maps of NGC 1976.

## 5. Summary

The map of the $^{13}$CO J=6-5 line, when combined with additional data, provides an estimate of the total CO column density that is more representative than those based on lines from lower rotational levels since the J=6 and J=5 energy levels have nearly the maximum population at kinetic temperatures of ~150 K and $H_2$ densities of $5 \cdot 10^6$ cm$^{-3}$. The prominence of the Hot Core is striking, especially when compared to lower rotational lines. The column density of CO in the nearby Orion Ridge is at least a factor of four smaller than in the Hot Core, although the column densities of $H_2$ are comparable. The CO/$H_2$ number ratio for the Hot Core is a factor of 5 lower than the usually quoted value, $10^{-4}$. The ratio for the Orion Ridge and Orion South are even lower. This may be due to chemical reactions at high temperatures. Since the intensity of the $^{13}$CO emission from PDRs should be about 0.01 of the intensity of the CO line, a comparison with the J=7-6 CO emission shows that at most <20% of the $^{13}$CO J=6-5 emission can arise in PDRs. Following Rodríguez-Franco et al. (2001), the emission from rotational lines of CN arises from the densest part of the PDR. Then the HII region NGC 1976 is to the east of the dense Orion Ridge. Further to the east there is less molecular material, so the expansion of the HII region in this direction is less hindered. From excitation considerations, the Hot Core must be located closer to the PDR than the Orion Ridge.

We thank M. Wolfire for advice about the interpretation of PDR emission in regard to $^{13}$CO. We also thank an anonymous referee for a careful and critical reading of the draft. J. K.'s work was carried out at the Jet Propulsion Laboratory, California Institute of Technology, under a contract with the National Aeronautics and Space Administration

*Facilities*: Arizona Radio Observatory (ARO) 10 meter Heinrich-Hertz-Telescope (HHT)

Zapata, L., Schmid-Burgk, J., Muders, D., Schilke, P., Menten, K., Guesten, R. 2010 A&A 510, A2

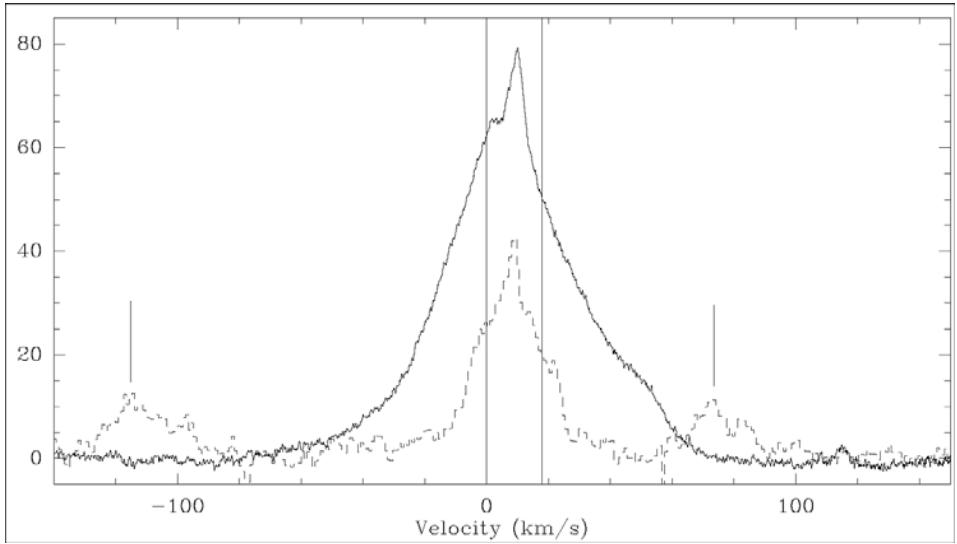

Fig. 1. (a) A comparison of our J=6-5 $^{13}$CO spectrum as a dashed line histogram, below, and the J=7-6 CO spectrum, above (Wilson et al. 2001b). Both were measured at the (0", 0") offset, at $\alpha$=05$^h$ 35$^m$ 14.4$^s$, $\delta$= -05° 22' 27" (J2000), the position of IRc2/I. The $^{13}$CO J=6-5 line frequency is 661.067 GHz. The vertical axis is corrected antenna temperature scale, $T_A^*$. The radial velocity scale is with respect to the local standard of rest, $V_{LSR}$. The two vertical lines, at 0 and 18 km s$^{-1}$, mark the gas that is not affected by an outflow. To mark additional emission features, we have drawn vertical lines to the sides of the J=6-5 $^{13}$CO spectrum. At -118 km s$^{-1}$ the feature is the sum of the (J, K)=(36,5) –(35,5) transition of methyl cyanide, the $J_k$=14$_2$ -13$_2$ transition of the $^{13}$C isotope of E –type methanol and the $J_{KaKc}$=21$_{7, 15}$ -21$_{6, 16}$ transition of sulfur dioxide. At +75 km s$^{-1}$ the feature is the $J_{KaKc}$=22$_{7, 15}$ to 22$_{6, 16}$ transition of sulfur dioxide.

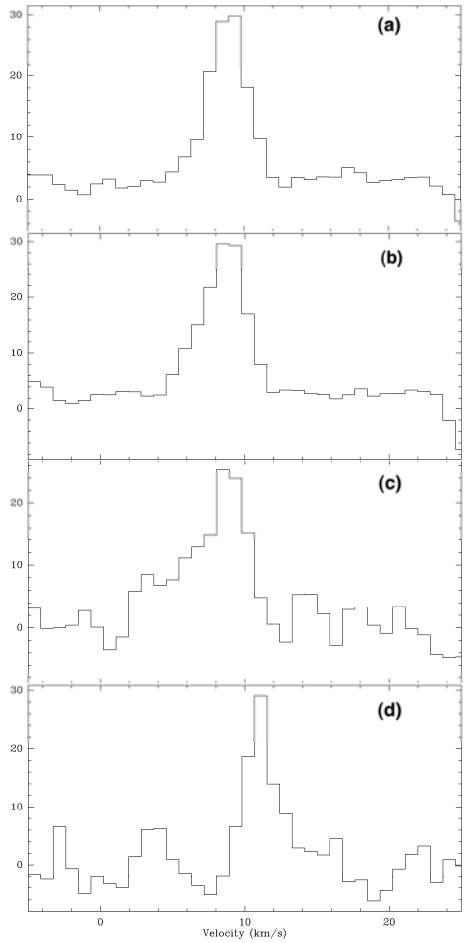

Fig. 2. Four spectra taken at positions offset from the zero point of our map, IRc2/I. The vertical axis is corrected antenna temperature scale, $T_A^*$. The radial velocity scale is with respect to the local standard of rest, $V_{LSR}$. (a) This spectrum was taken at the offset (0", -40"). (b) The spectrum was taken at the offset (0", -55"), which is at the Declination of $\theta^1$C Orionis, the brightest of the Trapezium stars. (c) This spectrum was taken at the offset (-24", -100"), which is the peak of the ammonia emission for the Orion South source (see, e.g., Zapata et al. 2010, Gaume et al. 1998). (d) This spectrum was taken at the offset (126", -146"), the location of the Orion Bar. Note that panel (c) has a slightly different vertical scale.

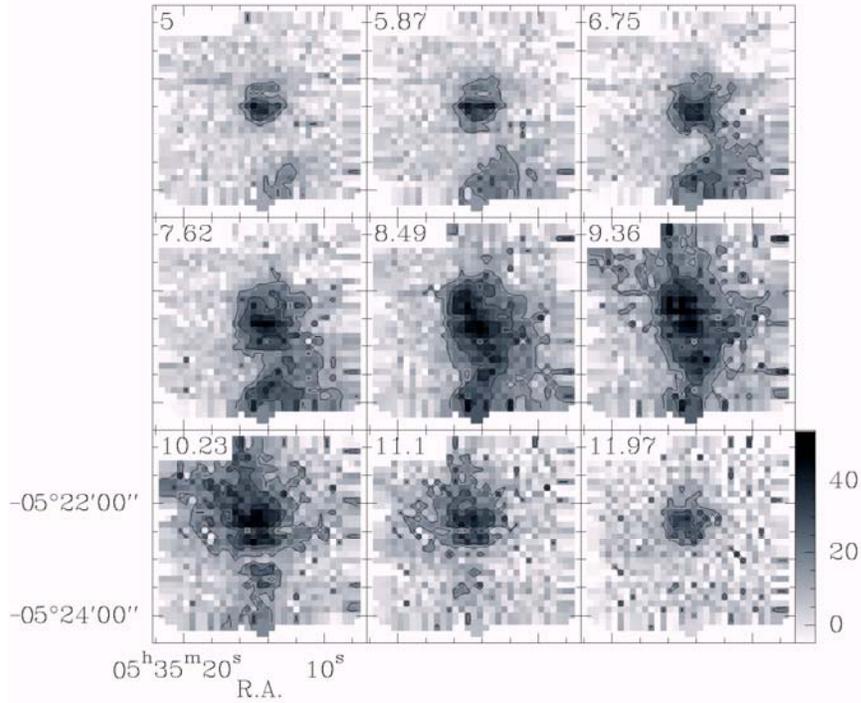

Fig. 3. Maps of a set of velocity channels in the J=6-5 $^{13}$CO emission line. The radial velocity of each channel is in the upper left of each panel. The FWHP beamsize is 13", and the contour steps are 10 K, $T_A^*$. The center position of the map is the position of the source IRc2/I. This is shown by a small cross. The coordinates are J2000.

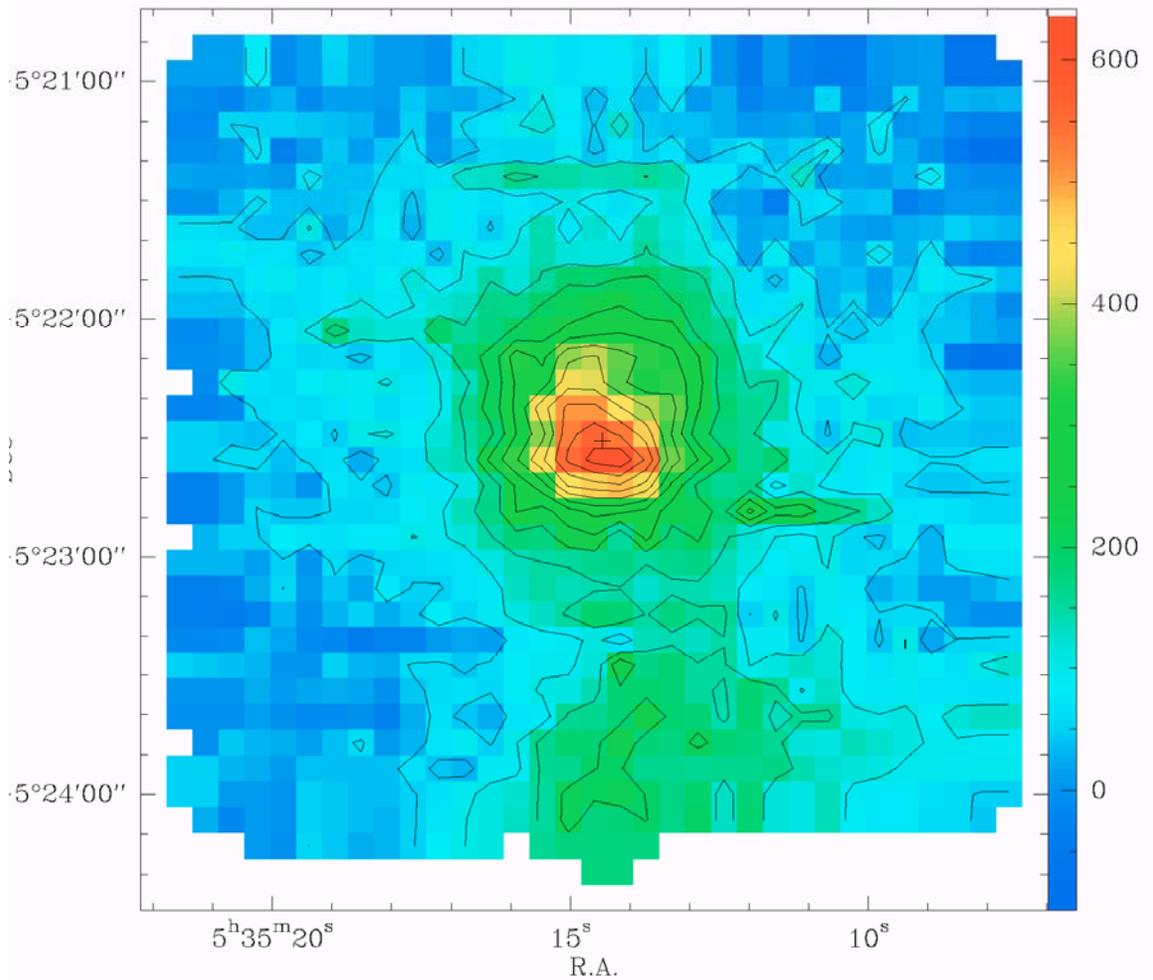

Fig. 4. A map of the J=6-5 $^{13}$CO emission integrated over the velocity range from 0 to 18 km s$^{-1}$. The contours are 50 K · km s$^{-1}$ to 600 K·km s$^{-1}$ in steps of 50 K· km s$^{-1}$, $T_A^*$ and the FWHP beamsize for this map is 13". The cross marks the reference position of IRc2/I. The coordinates are epoch J2000.

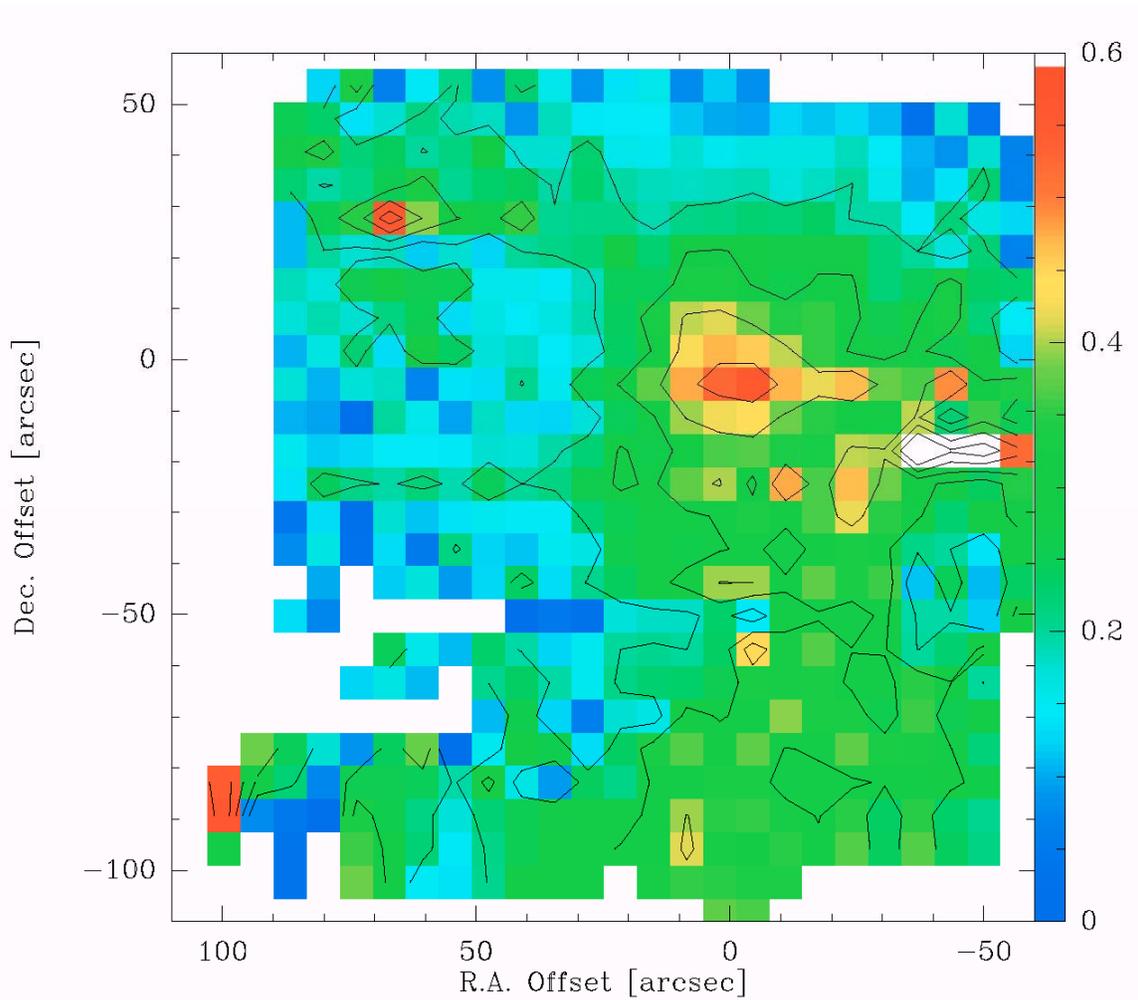

Figure 5. A map of the ratio of the $^{13}CO$ J=6-5 emission to the CO J=7-6 emission, both integrated from 0 to +18 km s$^{-1}$. See Figure 1 for the limits of integration on the line profile for the (0" 0") position. The ratio has a maximum at the position of the Hot Core (the (0" 0") position) and a compact minimum close to the Declination of the Trapezium star, $\theta^1$C Orionis, at $\Delta\delta$=-55".

Table 1: Column densities and ratios

| Source | $T_K$ (K) | CO Column Density $cm^{-2}$ | $H_2$ Column Density $cm^{-2}$ | Ratio CO to $H_2$ |
|---|---|---|---|---|
| Hot Core | 150 | $5.2\ 10^{19}$ | $3.0\ 10^{24}$ | $2\ 10^{-5}$ |
| Orion Ridge | 50 | $4.6\ 10^{18}$ | $1.1\ 10^{24}$ | $4\ 10^{-6}$ |
| Orion South | 100 | $4.3\ 10^{18}$ | $1.8\ 10^{24}$ | $2\ 10^{-6}$ |